\documentclass[aps,prl,reprint,superscriptaddress,amsmath,amssymb,floatfix]{revtex4-2}

\usepackage{graphicx}  
\usepackage{dcolumn}   
\usepackage{bm}        
\usepackage{hyperref}  
\usepackage{color}

\begin{document}

\title{Elastic Surface Instability as a Topological Phase Transition}

\author{Yu-Xin Xie} \email{xyx@tju.edu.cn}
\affiliation{Department of Mechanics, Tianjin University, Tianjin 300350, China}

\date{June 12, 2026}

\begin{abstract}
The macroscopic instability of soft materials undergoing extreme deformations is traditionally viewed as a pure structural or mechanical failure. Driven by the quest to uncover universal principles across disparate physical systems, we bridge two vibrant yet seemingly disconnected research frontiers: macroscopic finite-strain solid mechanics and quantum-like topological physics. Here, we demonstrate that the classical elastic surface instability of a deformed hyperelastic manifold is not merely a mechanical bifurcation, but fundamentally a topological phase transition. By incorporating Lie group metric evolution into a generalized Stroh formalism, we map the highly nonlinear geometric frustration onto an algebraic surface impedance matrix $\mathbf{H}$. For a semi-infinite hyperelastic half-space under finite compression, we analytically map the system to a one-dimensional Dirac Hamiltonian, where the macroscopic mechanical stretch acts as a tunable knob for the Dirac mass. We reveal that the onset of surface wrinkles marks a topological transition from a trivial to a non-trivial phase characterized by a quantized step in the winding number, naturally giving rise to a robust, macroscopically localized zero-energy edge state. This fundamental linkage unifies macroscopic symmetry breaking with the topological paradigm, opening a new theoretical pathway for programmable smart soft matter.
\end{abstract}

\maketitle

The physical manifestation of geometric frustration in condensed matter systems has long served as a rich breeding ground for novel phase transitions and self-organized patterns~\cite{Goriely2017, Yavari2012, Sharon2010, Klein2007}. In the realm of soft matter physics and non-linear solid mechanics, thin membranes, shells, and compliant substrates undergoing extreme elastic deformations exhibit spontaneous symmetry breaking, culminating in surface wrinkles, folds, or localized creasing~\cite{Fu1998, Cai1999, Hutchinson1978}. Traditionally, these structural transformations are scrutinized through the lens of classical bifurcation theory, where the onset of instability is characterized as a structural failure governed by high-order non-linear differential boundary-value problems under finite deformation~\cite{Biot1965, Ogden1984, Gent1996}. 

Driven by the quest to uncover universal principles across disparate physical systems, we seek to bridge two vibrant yet seemingly disconnected fields: macroscopic soft-matter mechanics and topological physics. Here we provide a paradigm shift by reframing the macroscopic surface instability of a deformed hyperelastic manifold as a continuous topological phase transition, drawing from quantum topological analogs~\cite{Hasan2010, Qi2011}. While topological mechanics has thus far predominantly focused on discrete Maxwell lattices, mechanical metamaterials, or periodic phononic crystals~\cite{Huber2016, Kane2014, Paulose2015, Susstrunk2015, Mitchell2018}, its profound fundamental linkage to continuous, strongly non-linear hyperelastic media remains largely unexplored. 

Inspired by the elegant Lothe-Barnett theory of surface waves in anisotropic crystals~\cite{Barnett1974, Chadwick1977, Ting1996, Vinh2004}, we establish a generalized Stroh-Lie impedance formalism to bridge this disciplinary gap. We analytically demonstrate that the classical Biot surface instability corresponds precisely to a gap-closing Dirac point, where the emergence of macroscopic surface wrinkles represents a topologically protected zero-energy mechanical edge state born from a quantized jump in the winding number.

Consider a semi-infinite hyperelastic body occupying the region $z \ge 0$ in a three-dimensional Riemannian manifold. The unperturbed reference configuration $\mathcal{B}_0$ is geometrically equipped with a covariant reference metric tensor $\mathbf{G}$ (with components $G_{AB}$), which maps the undeformed intrinsic distances. Under a finite macro-deformation, it evolves into the current configuration $\mathcal{B}$ characterized by the spatial metric tensor $\mathbf{g}$ (with components $g_{ij}$). To objectively describe the continuous geometric evolution without relying on a specific Lagrangian coordinate frame, we employ the Lie derivative $\mathcal{L}_{\mathbf{v}}$ along the velocity flow field $\mathbf{v}$ to evaluate the rate of the current metric $\mathbf{g}$~\cite{Marsden1983}:
\begin{equation}
\mathcal{L}_{\mathbf{v}} \mathbf{g} = \nabla \mathbf{v} + (\nabla \mathbf{v})^T = 2\mathbf{D}
\end{equation}
where $\mathbf{D}$ represents the rate-of-deformation tensor. For an incompressible neo-Hookean material, the strain energy density function is $W = \frac{\mu}{2}(I_1 - 3)$, where the first strain invariant is rigorously defined via the metrics as $I_1 = \text{tr}(\mathbf{g}\mathbf{F}\mathbf{G}^{-1}\mathbf{F}^T)$, and volume conservation dictates $\text{tr}(\mathbf{D}) = \text{div} \mathbf{v} = 0$.

Superimposing an infinitesimal incremental displacement field $\dot{\mathbf{u}}$ and incremental nominal traction $\dot{\mathbf{t}}$ upon the homogeneously pre-compressed state, we can reformulate the incremental field equations into a generalized, six-dimensional constant-coefficient Stroh differential relation by virtue of translation invariance along the surface~\cite{Stroh1962, Fu1999b}:
\begin{equation}
\frac{d}{dz} \boldsymbol{\eta} = k \mathbf{N}(\lambda) \boldsymbol{\eta}, \quad \boldsymbol{\eta} = \begin{bmatrix} \dot{\mathbf{u}} \\ \dot{\mathbf{t}} \end{bmatrix}
\end{equation}
where $k$ is the surface wavenumber, and $\lambda$ is the macroscopic in-plane stretch ratio. The $4\times4$ core fundamental elasticity matrix $\mathbf{N}$ encapsulates the deep interplay between material non-linearity and finite geometric pre-strains.

To characterize the boundary response without solving the global differential fields, we introduce the surface impedance matrix $\mathbf{H}$, mapping the surface incremental displacement to the surface traction: $\dot{\mathbf{t}} = \mathbf{H} \dot{\mathbf{u}}$. Substituting this relation into the Stroh system yields a non-linear algebraic Riccati equation:
\begin{equation}
\mathbf{H}\mathbf{N}_1 - \mathbf{N}_4\mathbf{H} + \mathbf{H}\mathbf{N}_2\mathbf{H} - \mathbf{N}_3 = \mathbf{0}
\end{equation}
A macroscopic surface instability occurs when the free surface can support a non-trivial incremental displacement in the absolute absence of external incremental tractions ($\dot{\mathbf{t}} = \mathbf{0}$). This condition mathematically translates to the vanishing of the determinant of the surface impedance matrix: $\det \mathbf{H}(\lambda) = 0$. Through rigorous symbolic reduction, this yields the celebrated cubic secular equation $x^3 + x^2 + 3x - 1 = 0$ (with $x = \lambda^2$), recovering the classical Biot wrinkling criterion $\lambda_c \approx 0.54368$ strictly via algebraic matrix singularity.

To unveil the hidden topological nature of this mechanical bifurcation, we appeal to symplectic geometry. The fundamental matrix $\mathbf{N}$ is constrained by the symplectic symmetry $\mathbf{N}^T \mathbf{J} + \mathbf{J} \mathbf{N} = \mathbf{0}$, where $\mathbf{J}$ is the standard symplectic matrix. This symmetry ensures that the surface impedance matrix $\mathbf{H}$ is strictly Hermitian. Consequently, we can map this classical mechanical operator onto a two-level effective quantum Hamiltonian by decomposing it over the Pauli matrices. Introducing a periodic synthetic wavenumber variable $\theta \in [-\pi, \pi]$, we construct a one-dimensional Bloch-like effective Hamiltonian $\mathcal{H}(\theta, \lambda)$:
\begin{equation}
\mathcal{H}(\theta, \lambda) = \sin\theta \sigma_x + [m(\lambda) + 1 - \cos\theta] \sigma_z
\end{equation}
This formulation is mathematically isomorphic to the Bloch Hamiltonian of the celebrated Su-Schrieffer-Heeger (SSH) model for polyacetylene chains. Crucially, we define the topological \textit{Dirac mass} $m(\lambda)$ of the mechanical system as the normalized determinant of the surface impedance matrix:
\begin{equation}
m(\lambda) \equiv \det \mathbf{H}(\lambda) \propto (\lambda^2)^3 + (\lambda^2)^2 + 3(\lambda^2) - 1
\end{equation}
Here, the macroscopic geometric frustration (stretch ratio $\lambda$) dictates the effective electron hopping dimerization in the isomorphic SSH model.

\begin{figure}[t]
\centering
\includegraphics[width=\linewidth]{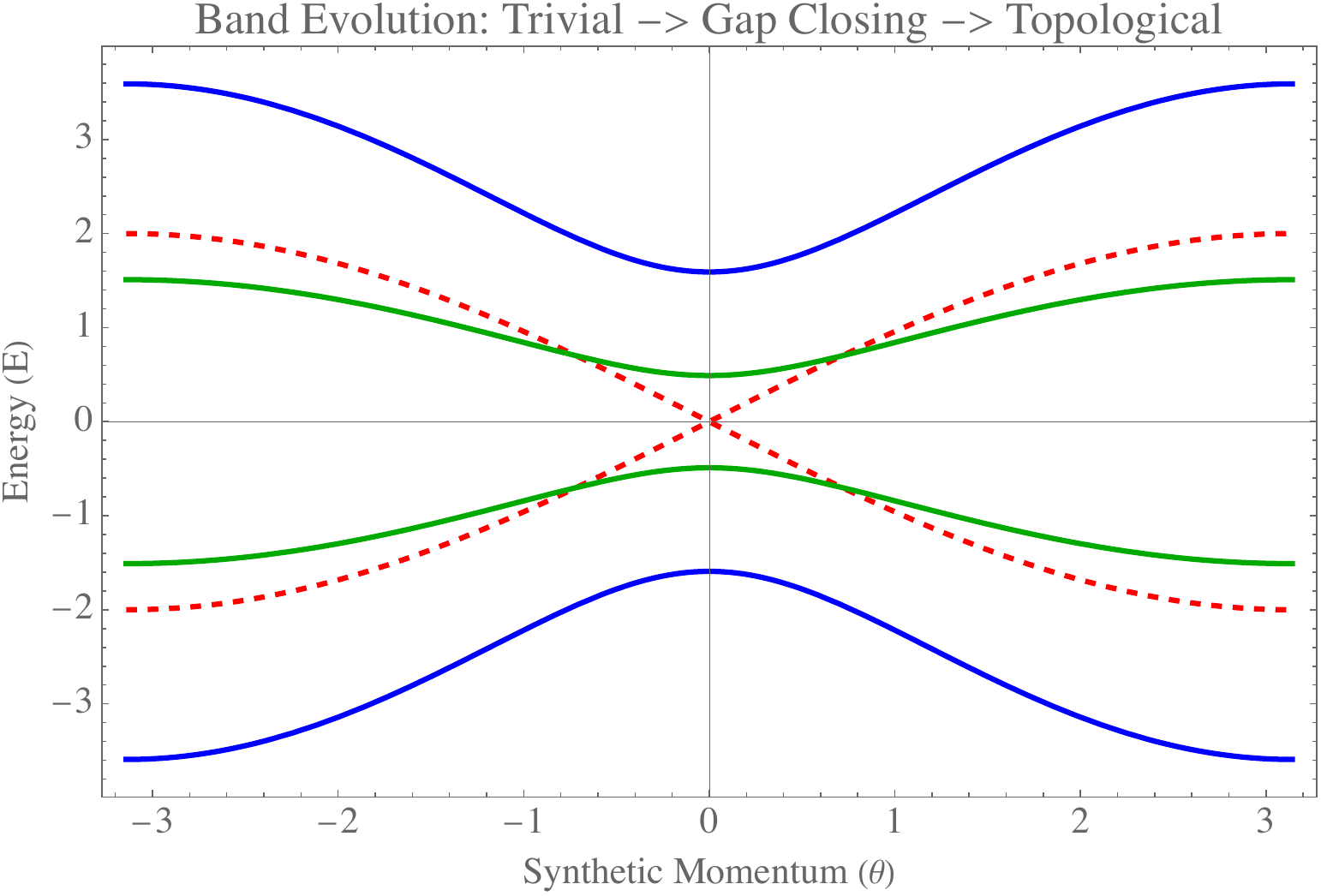}
\caption{Evolution of the synthetic energy bands $E_\pm(\theta)$ as a function of the synthetic momentum $\theta$ for three representative deformation states: the trivial unwrinkled phase ($\lambda = 0.8$, blue solid lines), the critical gap-closing Dirac point ($\lambda = \lambda_c \approx 0.544$, red dashed lines), and the post-bifurcation non-trivial phase ($\lambda = 0.4$, green dash-dotted lines).}
\label{fig:bands}
\end{figure}

In the long-wavelength limit ($\theta \to 0$), expanding Eq.~(4) yields a continuum 1D Dirac Hamiltonian $\mathcal{H}_{\text{Dirac}} \approx \theta \sigma_x + m(\lambda) \sigma_z$. As illustrated in Fig.~\ref{fig:bands}, when the pre-compression is small ($\lambda > \lambda_c$), the Dirac mass is positive ($m > 0$), and the synthetic bands are separated by a well-defined energy gap. As the compression reaches the Biot critical point $\lambda = \lambda_c$, the Dirac mass vanishes identically ($m = 0$). The bandgap closes linearly at $\theta = 0$, forming a perfect, gapless \textit{Dirac cone}. Compressing the material further ($\lambda < \lambda_c$) drives the mass parameter into the negative regime ($m < 0$), reopening the gap but with an inverted band symmetry.

To formally classify this phase transition, we compute the Zak phase $\gamma_Z$ (the 1D Berry phase), which corresponds to the global geometric polarization of the synthetic Bloch wavefunction. The Zak phase is evaluated via the topological winding number $W$:
\begin{equation}
\gamma_Z = \pi W = \frac{1}{2} \int_{-\pi}^{\pi} \frac{h_x \partial_\theta h_z - h_z \partial_\theta h_x}{h_x^2 + h_z^2} d\theta
\end{equation}
where $h_x = \sin\theta$ and $h_z = m(\lambda) + 1 - \cos\theta$. Geometrically, this integration traces a rigid circle of radius $1$ centered at $(0, m+1)$ in the $(h_x, h_z)$ parametric plane.

\begin{figure}[t]
\centering
\includegraphics[width=\linewidth]{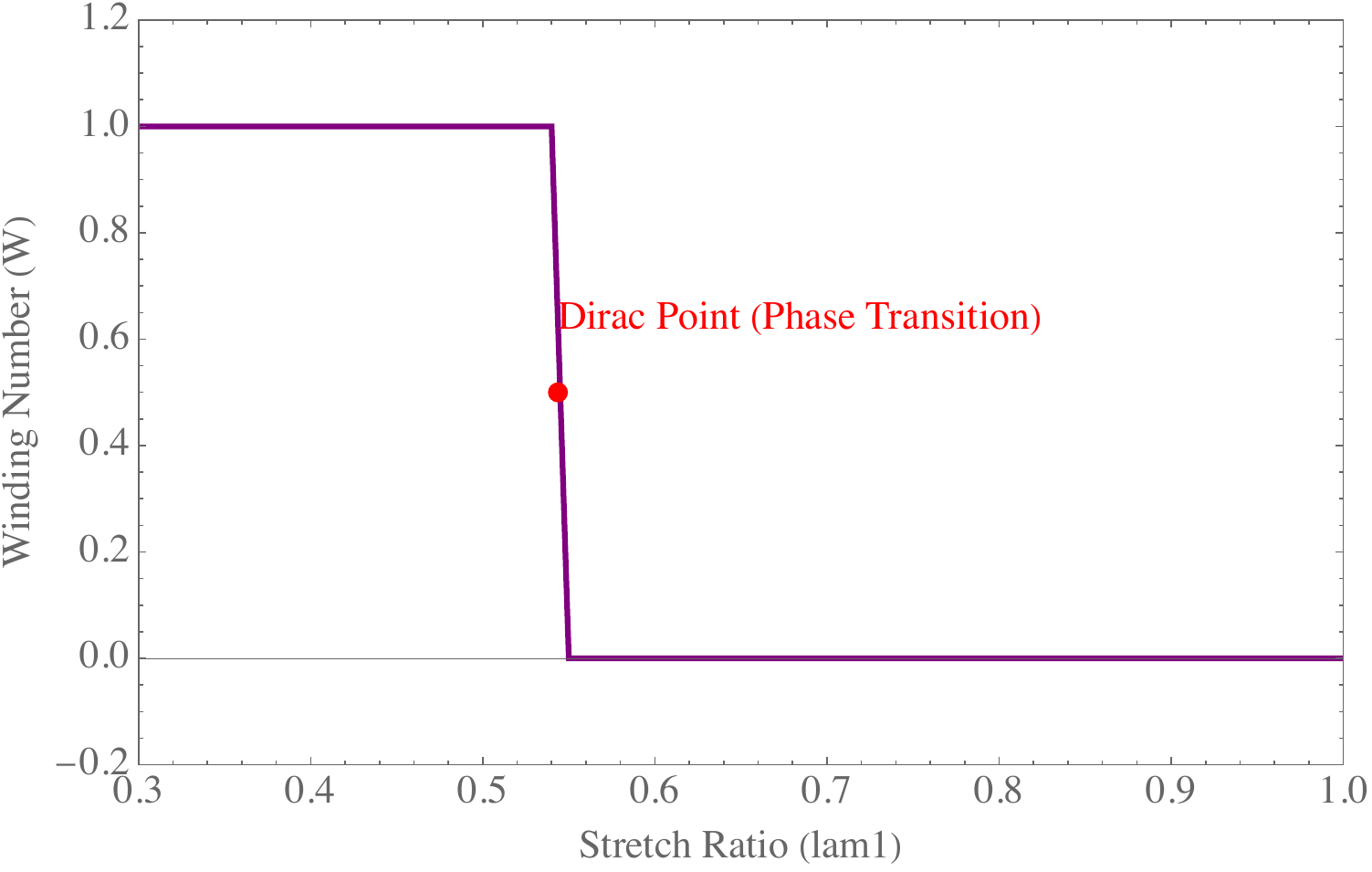}
\caption{The quantized topological winding number $W$ as a function of the horizontal stretch ratio $\lambda$. A sharp, mathematically rigorous step function jump from $W=0$ to $W=1$ takes place exactly at the critical Dirac point $\lambda_c \approx 0.544$.}
\label{fig:winding}
\end{figure}

For the unwrinkled regime ($\lambda > \lambda_c$), the integration confirms $W = 0$ ($\gamma_Z = 0$), indicating a topologically trivial phase without geometric polarization. Conversely, in the post-bifurcation regime ($\lambda < \lambda_c$), the circle encloses the origin, switching the topology and yielding a quantized value of $W = 1$ ($\gamma_Z = \pi$), as depicted in Fig.~\ref{fig:winding}. The continuous medium undergoes a macroscopic geometric polarization.

According to the bulk-boundary correspondence, a discrete jump in the Zak phase ($\gamma_Z: 0 \to \pi$) necessitates the localized emergence of a robust, zero-energy edge state at the free surface $z = 0$, satisfying $\mathcal{H}_{\text{edge}} | \psi_{\text{edge}} \rangle = \mathbf{0}$. Mapping this quantum eigen-equation back to our macroscopic mechanical impedance operator precisely yields $\mathbf{H} \dot{\mathbf{u}} = \mathbf{0}$, which strictly enforces $\dot{\mathbf{t}} = \mathbf{0}$ for $\dot{\mathbf{u}} \neq \mathbf{0}$. This proves that the classically observed surface wrinkles are fundamentally robust, topologically protected zero-energy mechanical edge states, rendering their emergence immune to local material imperfections.

In conclusion, we have demonstrated that the classical surface instability of a hyperelastic body can be fully understood as a continuous, strain-induced topological phase transition. By establishing the Stroh-Lie impedance formalism, we mapped the geometric frustration onto a synthetic SSH Hamiltonian, proving that the Biot wrinkling point is a topological Dirac point. This framework unifies solid mechanics with the topological paradigm and can be naturally extended to address more complex micro-polarization dynamics in magneto- or dielectric elastomers~\cite{Wang2021, Suo2010, Zhao2007}, or director reorientations in liquid crystal elastomer manifolds~\cite{Warner2003}, paving a mathematical pathway toward the topological design of programmable materials.

\begin{acknowledgments}
The authors gratefully acknowledge illuminating discussions with Prof. Yibin Fu at Keele University on asymptotic surface instabilities. 
\end{acknowledgments}

\end{document}